**Title:**

A fast quantum interface between different spin qubit encodings

**Authors:**

A. Noiri[1†*], T. Nakajima[1†], J. Yoneda[1], M. R. Delbecq[1,2], P. Stano[1], T. Otsuka[1,3,4], K. Takeda[1], S. Amaha[1], G. Allison[1], K. Kawasaki[5], Y. Kojima[5], A. Ludwig[6], A. D. Wieck[6], D. Loss[1,7], and S. Tarucha[1,5*]

**Affiliations:**

[1]*RIKEN, Center for Emergent Matter Science (CEMS), Wako-shi, Saitama 351-0198, Japan*

[2]*Laboratoire Pierre Aigrain, Ecole Normale Supérieure-PSL Research University, CNRS, Université Pierre et Marie Curie-Sorbonne Universités, Université Paris Diderot-Sorbonne Paris Cité, 24 rue Lhomond, 75231 Paris Cedex 05, France*

[3]*Research Institute of Electrical Communication, Tohoku University, 2-1-1 Katahira, Aoba-ku, Sendai 980-8577, Japan*

[4]*JST, PRESTO, 4-1-8 Honcho, Kawaguchi, Saitama 332-0012, Japan*

[5]*Department of Applied Physics, University of Tokyo, 7-3-1 Hongo, Bunkyo-ku, Tokyo 113-8656, Japan*

[6]*Lehrstuhl für Angewandte Festkörperphysik, Ruhr-Universität Bochum, D-44780 Bochum, Germany*

[7]*Department of Physics, University of Basel, Klingelbergstrasse 82, 4056 Basel, Switzerland*

[†]These authors contributed equally to this work.

[*]e-mail: akito.noiri@riken.jp or tarucha@ap.t.u-tokyo.ac.jp

**Single-spin qubits in semiconductor quantum dots proposed by Loss and DiVincenzo (LD qubits) hold promise for universal quantum computation with demonstrations of a high single-qubit gate fidelity above 99.9 % and two-qubit gates in conjunction with a long coherence time. However, initialization and readout of a qubit is orders of magnitude slower than control, which is detrimental for implementing measurement-based protocols such as error-correcting codes. In contrast, a singlet-triplet (ST) qubit, encoded in a two-spin subspace, has the virtue of fast readout with high fidelity and tunable coupling to the electric field. Here, we present a hybrid system which benefits from the different advantages of these two distinct spin-qubit implementations. A quantum interface between the two codes is realized by electrically tunable inter-qubit exchange coupling. We demonstrate a controlled-phase (CPHASE) gate that acts within 5.5 ns, much faster than the measured dephasing time of 211 ns. The presented hybrid architecture will be useful to settle remaining key problems with building scalable spin-based quantum computers.**

Initialization, single- and two-qubit gate operations, and measurements are fundamental elements for universal quantum computation[1]. Generally, they should all be fast and with high fidelity to reach the fault-tolerance thresholds[2]. So far, various encodings of spin qubits into one to three-spin subspaces have been developed in semiconductor quantum dots[3-15]. In particular, recent experiments demonstrated all of these elements including two-qubit logic gates for LD and ST qubits[6-8,14]. These qubits have different advantages depending on the gate operations, and combinations thereof can increase the performance of spin-based quantum computing. In LD qubits, the two-qubit gate is fast[6,7] as it relies on the exchange interaction between neighboring spins. In contrast, the two-qubit gate in ST qubits is much slower as it is mediated by a weak dipole coupling[14]. Concerning initialization and readout, however, the situation is the opposite: it is slow for LD qubits, relying on spin-selective tunneling to a lead[16,17], while it is orders of magnitude faster in ST qubits relying on Pauli spin blockade[12,13]. Therefore, a fast and reliable interface between LD and ST qubits would allow for merging the advantages of both realizations. In this work, we demonstrate such an interface that implements a CPHASE gate based on nearest neighbor exchange coupling in a quantum dot array[18,19].

**Results:**

**Demonstration of a LD qubit and a ST qubit formed in a triple quantum dot.**

A hybrid system comprising a LD qubit and a ST qubit is implemented in a linearly-coupled gate-defined triple quantum dot (TQD) shown in Fig. 1a. The LD qubit ($Q_{LD}$) is formed in the left dot while the ST qubit ($Q_{ST}$) is hosted in the other two dots. We place a micro-magnet near the TQD to coherently control $Q_{LD}$ via electric dipole spin resonance (EDSR)[20-23,26]. At the same time it makes the Zeeman energy difference between the center and right dots, $\Delta E_Z^{ST}$, much larger than their exchange coupling $J^{ST}$, such that the eigenstates of $Q_{ST}$ become $|\uparrow\downarrow\rangle$ and $|\downarrow\uparrow\rangle$ rather than singlet $|S\rangle$ and triplet $|T\rangle$. We apply an external in-plane magnetic field $B_{ext} = 3.166$ T to split the $Q_{LD}$ states by the Zeeman energy $E_Z$ as well as to separate polarized triplet states $|\uparrow\uparrow\rangle$ and $|\downarrow\downarrow\rangle$ from the $Q_{ST}$ computational states. The experiment is conducted in a dilution refrigerator with an electron temperature of approximately 120 mK. The qubits are manipulated in the $(N_L, N_C, N_R) = (1,1,1)$ charge state while the $(1,0,1)$ and $(1,0,2)$ charge states are also used for initialization and readout (see Fig. 1b). Here, $N_{L(C,R)}$ denotes the number of electrons inside the left (center, right) dot.

We first independently measure the coherent time evolution of each qubit to calibrate the initialization, control and readout. We quench the inter-qubit exchange coupling by largely detuning the energies of the $(1,1,1)$ and $(2,0,1)$ charge states. For $Q_{LD}$, we observe Rabi oscillations[4] with a frequency $f_{Rabi}$ of up to 10 MHz (Fig. 1d) as a function of the microwave (MW) burst time $t_{MW}$, using the pulse sequence in Fig. 1e. For $Q_{ST}$, we observe the precession between $|S\rangle$ and $|T\rangle$ (ST precession) (Fig. 1f) as a function of the evolution time $t_e$, using the pulse sequence in Fig. 1g (see

Supplementary Material, Sec. 2 for full control of Q$_{ST}$). We use a metastable state to measure Q$_{ST}$ with high fidelity[13] (projecting to $|S\rangle$ or $|T\rangle$) in the presence of large $\Delta E_Z^{ST}$ with which the lifetime of $|T\rangle$ is short[27].

**The two-qubit coupling calibration.**

The two qubits are interfaced by exchange coupling $J^{QQ}$ between the left and center dots as illustrated in Fig. 1c. We operate the two-qubit system under the conditions of $E_Z \gg \Delta E_Z^{ST}, \Delta E_Z^{QQ} \gg J^{QQ} \gg J^{ST}$ where $\Delta E_Z^{QQ}$ is the Zeeman energy difference between the left and center dots. Then, the Hamiltonian of the system is

$$\mathcal{H} = -E_Z \hat{\sigma}_z^{LD}/2 - \Delta E_Z^{ST} \hat{\sigma}_z^{ST}/2 + J^{QQ}(\hat{\sigma}_z^{LD}\hat{\sigma}_z^{ST} - 1)/4 \qquad (1)$$

where $\hat{\sigma}_z^{LD}$ and $\hat{\sigma}_z^{ST}$ are the Pauli z-operators of Q$_{LD}$ and Q$_{ST}$, respectively[18] (Supplementary Material, Sec. 3). The last term in Eq. (1) reflects the effect of the inter-qubit coupling $J^{QQ}$: for states in which the spins in the left and center dots are antiparallel, the energy decreases by $J^{QQ}/2$ (see Fig. 2a). In the present work, we choose to operate Q$_{LD}$ as a control qubit and Q$_{ST}$ as a target, although these are exchangeable. With this interpretation, the ST precession frequency $f^{ST}$ depends on the state of Q$_{LD}$, $f_{\sigma_z^{LD}}^{ST} = (\Delta E_Z^{ST} - \sigma_z^{LD} J^{QQ}/2)/h$. Here $\sigma_z^{LD}$ represents $|\uparrow\rangle$ or $|\downarrow\rangle$ and +1 or -1 interchangeably. This means that while $J^{QQ}$ is turned on for the interaction time $t_{int}$, Q$_{ST}$ accumulates the controlled-phase $\phi_C = 2\pi J^{QQ} t_{int}/h$, which provides the CPHASE gate (up to single-qubit phase gates; see Supplementary Material, Sec. 7) in $t_{int} = h/2J^{QQ}$. An important feature of this two-qubit gate is that it is intrinsically fast, scaling with $J^{QQ}/h$ which can be tuned up to ~100 MHz, and is limited only by the requirement $J^{QQ} \ll \Delta E_Z^{QQ}$ ~500 MHz in our device.

Before testing the two-qubit gate operations, we calibrate the inter-qubit coupling strength $J^{QQ}$, and its tunability by gate voltages. The inter-qubit coupling in pulse stage F (Fig. 2b) is controlled by the detuning energy between (2,0,1) and (1,1,1) charge states (one of the points denoted E in Fig. 1b). To prevent leakage from the Q$_{ST}$ computational states, we switch $J^{QQ}$ on and off adiabatically with respect to $\Delta E_Z^{QQ}$ by inserting voltage ramps to stage F with a total ramp time of $t_{ramp} = 24$ ns (Fig. 2b)[28]. The coherent precession of Q$_{ST}$ is measured by repeating the pulse stages from D to H without initializing, controlling and measuring Q$_{LD}$, which makes Q$_{LD}$ a random mixture of $|\uparrow\rangle$ and $|\downarrow\rangle$. Fig. 2c shows the FFT spectra of the precession measured for various interaction points indicated in Fig. 1b. As we bring the interaction point closer to the boundary of (1,1,1) and (2,0,1), $J^{QQ}$ becomes larger and we start to see splitting of the spectral peaks into two. The separation of the two peaks is given by $J^{QQ}/h$ which can be controlled by the gate voltage as shown in Fig. 2d.

We now demonstrate the controllability of the ST precession frequency by the input state of Q$_{LD}$, the essence of a CPHASE gate. We use the quantum circuit shown in Fig. 2b, which combines the pulse sequences for independent characterization of $Q_{LD}$ and $Q_{ST}$. Here we choose the interaction point such that $J^{QQ}/h = 90$ MHz. By using either $|\uparrow\rangle$ or $|\downarrow\rangle$ as the $Q_{LD}$ initial state (the latter prepared by an EDSR $\pi$ pulse), we observe the ST precessions as shown in Fig. 2e. The data fit well to Gaussian-decaying oscillations giving $f_{|\uparrow\rangle}^{ST} = 434 \pm 0.5$ MHz and $f_{|\downarrow\rangle}^{ST} = 524 \pm 0.4$ MHz [These are consistent with the values determined by Bayesian estimation discussed in Methods]. This demonstrates the control of the precession rate of $Q_{ST}$ by $J^{QQ}/h$ depending on the state of $Q_{LD}$.

**Demonstration of a CPHASE gate.**

To characterize the controlled-phase accumulated during the pulse stage F, we separate the phase of $Q_{ST}$ into controlled and single-qubit contributions as $\phi_{\sigma_z^{LD}} = -\pi\sigma_z^{LD}J^{QQ}(t_{int} + t_0)/h$ and $\phi^{ST} = 2\pi\Delta E_Z^{ST}(t_{int} + t_{ramp})/h + \phi_0$, respectively. Here $t_0(\ll t_{ramp})$ represents the effective time for switching on and off $J^{QQ}$ (Supplementary Material, Sec. 5). A phase offset $\phi_0$ denotes the correction accounting for nonuniform $\Delta E_Z^{ST}$ during the ramp (Supplementary Material, Sec. 5). Then the probability of finding the final state of $Q_{ST}$ in singlet is modeled as

$$P_{S,model} = a\cos(\phi_{\sigma_z^{LD}} + \phi^{ST})\exp(-(t_{int}/T_2^*)^2) + b \qquad (2)$$

where $a$, $b$ and $T_2^*$ represent the values of amplitude, mean and the dephasing time of the ST precession, respectively. We use maximum likelihood estimation (MLE) combined with Bayesian estimation[29,30] to fit all variables in Eq. (2), that are $a, b, t_0, J^{QQ}, T_2^*, \phi_0$, and $\Delta E_Z^{ST}$, from the data (Methods). This allows us to extract the $t_{int}$ dependence of $\phi_{\sigma_z^{LD}}$ (Fig. 3a) (Methods) and consequently $\phi_C = \phi_{|\downarrow\rangle} - \phi_{|\uparrow\rangle}$ (Fig. 3b). It evolves with $t_{int}$ in the frequency of $J^{QQ}/h = 90$ MHz, indicating that the CPHASE gate time can be as short as $h/J^{QQ} = 5.5$ ns (up to single-qubit phase). On the other hand, $T_2^*$ obtained in the MLE is 211 ns, much longer than what is observed in Fig. 2e because the shorter data acquisition time used here cuts off the low-frequency component of the noise spectrum[29]. We note that this $T_2^*$ is that for the two-qubit gate while $J^{QQ}$ is turned on[8], and therefore it is likely to be dominated by charge noise rather than the nuclear field fluctuation (Supplementary Material, Sec. 6). The ratio $2J^{QQ}T_2^*/h$ suggests that 38 CPHASE operations would be possible within the two-qubit dephasing time. We anticipate that this ratio can be further enhanced by adopting approaches used to reduce the sensitivity to charge noise in exchange gates such as symmetric operation[31,32] and operation in an enhanced field gradient[33].

Finally we show that the CPHASE gate operates correctly for arbitrary $Q_{LD}$ input states. We implement the circuit shown in Fig. 4a in which $t_{int}$ is fixed to yield $\phi_C = \pi$, while a coherent intitial $Q_{LD}$ state with an arbitrary $\sigma_z^{LD}$ is prepared by EDSR. We extract the averaged $\phi_{\sigma_z^{LD}}$, $\langle\phi_{\sigma_z^{LD}}\rangle$ by Bayesian estimation[29,30], which shows an oscillation as a function of $t_{MW}$ in agreement with the Rabi oscillation measured independently by reading out $Q_{LD}$ at stage C as shown in Fig. 4b (see Methods for the estimation procedure and the origin of the low visibility, i.e., $\max|\langle\phi_{\sigma_z^{LD}}\rangle| < \pi/2$). These results clearly demonstrate the CPHASE gate functioning for an arbitrary $Q_{LD}$ input state.

**Discussion:**

In summary, we have realized a fast quantum interface between a LD qubit and a ST qubit using a TQD. The CPHASE gate between these qubits is performed in 5.5 ns, much faster than its dephasing time of 211 ns and those ratio (∼38) would be high enough to provide a high fidelity CPHASE gate (Supplementary Material Sec. 8). Optimizing the magnet design to enhance the field gradient would allow even faster gate time beyond GHz with larger $J^{QQ}$. At the same time, this technique is directly applicable to Si-based devices with much better single-qubit coherence[5-9]. Our results suggest that the performance of certain quantum computational tasks can be enhanced by adopting different kinds of qubits for different roles. For instance, LD qubits can be used for high-fidelity control and long memory and the ST qubit for fast initialization and readout. This combination is ideal for example, the surface code quantum error correction where a data qubit must maintain the coherence while a syndrome qubit must be measured quickly[34]. Furthermore, the fast (~ 100 ns[25]) ST qubit readout will allow the read out of a LD qubit in a quantum-non-demolition manner[35] with a speed three orders of magnitude faster than a typical energy-selective tunneling measurement[16,17]. Viewed from the opposite side, we envisage coupling two ST qubits through an intermediate LD qubit, which would boost the two ST qubit gate speed by orders of magnitude compared to the demonstrated capacitive coupling scheme[14]. In addition, our results experimentally support the concept of the theoretical proposal of a fast two-qubit gate between two ST qubits based on direct exchange[36] which shares the same working principle as our two-qubit gate. Our approach will further push the demonstrated scalabitlity of spin qubits in quantum dot arrays beyond the conventional framework based on a unique spin-qubit encoding.

**Acknowledgement:**

Part of this work is financially supported by the ImPACT Program of Council for Science, Technology and Innovation (Cabinet Office, Government of Japan) the Grant-in-Aid for Scientific Research (No. 26220710), CREST (JPMJCR15N2, JPMJCR1675), JST, Incentive Research Project from RIKEN. AN acknowledges support from Advanced Leading Graduate Course for Photon Science (ALPS). TN acknowledges financial support from JSPS KAKENHI Grant Number 25790006. TO acknowledges financial support from Grants-in-Aid for Scientific Research (No. 16H00817, 17H05187), PRESTO (JPMJPR16N3), JST, Advanced Technology Institute Research Grant, the Murata Science Foundation Research Grant, Izumi Science and Technology Foundation Research Grant, TEPCO Memorial Foundation Research Grant, The Thermal & Electric Energy Technology Foundation Research Grant, The Telecommunications Advancement Foundation Research Grant, Futaba Electronics Memorial Foundation Research Grant, MST Foundation Research Grant, Kato Foundation for Promotion of Science Research Grant. A.D.W. and A.L. acknowledge gratefully support of DFG-TRR160, BMBF - Q.com-H 16KIS0109, and the DFH/UFA CDFA-05-06.

**Author contributions**

A.N. and J.Y. conceived the experiment. A.N. and T.N. performed the measurement with the assistance of K.K. and Y.K.. A.N. and T.N. conducted data analysis with the inputs from J.Y. and P.S.. A.N. and T.N. fabricated the device on the heterosutucture grown by A.L. and A.D.W.. A.N. and T.N. wrote the manuscript with inputs from other authors. All authors discussed the results. The project was supervised by S.T..

**Competing financial interests**

The authors declare that they have no competing financial interests.

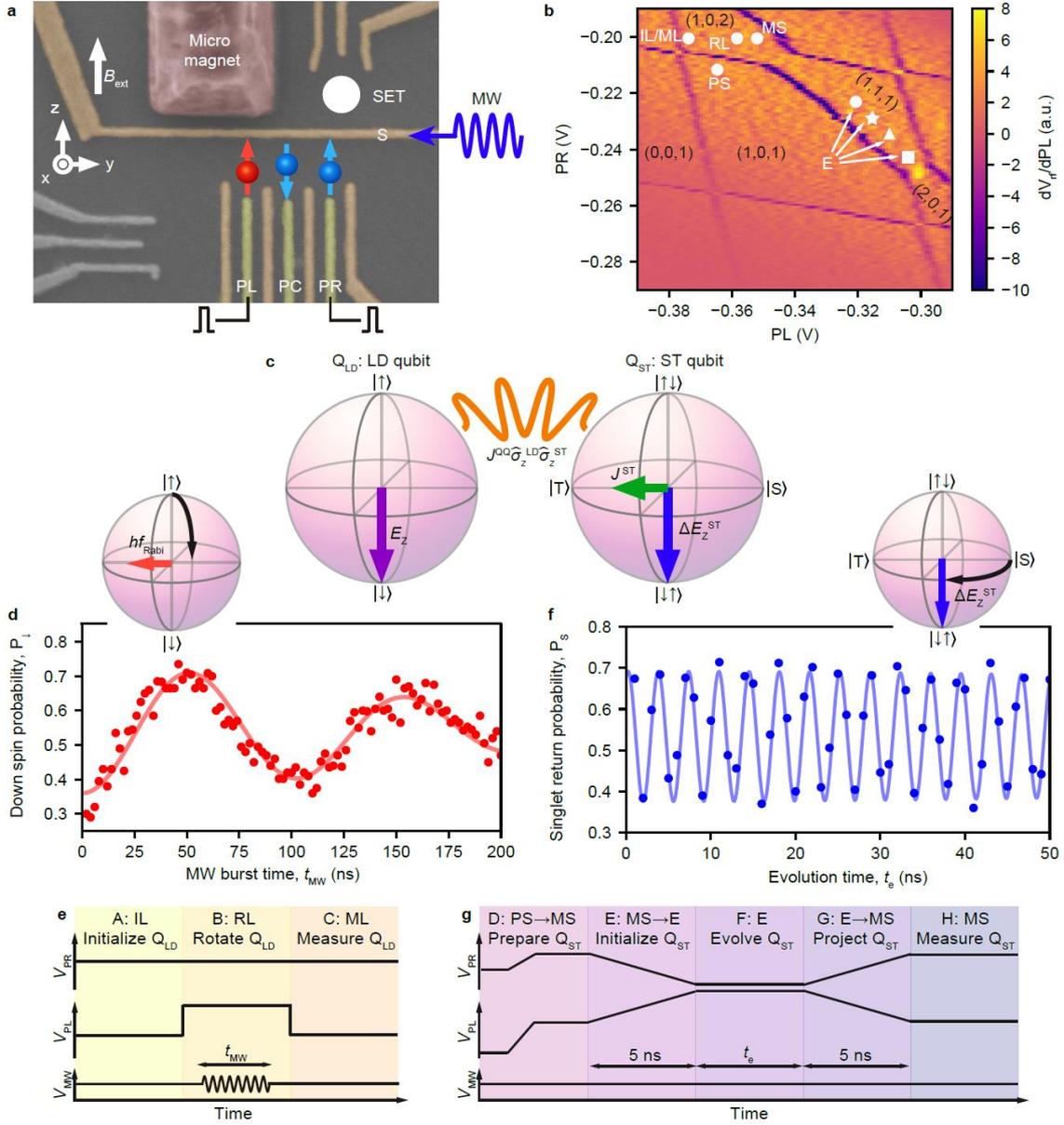

**Figure 1 | Hybrid system of a LD qubit and a ST qubit realized in a TQD.**

**a**, False colour scanning electron microscope image of a device identical to the one used in this study. The TQD is defined in a two-dimentional electron gas at the GaAs/AlGaAs heterointerface 100 nm below the surface. The upper single electron transistor is used for radiofrequency-detected charge sensing[24,25]. A MW with a frequency of 17.26 GHz is applied to the S gate to drive EDSR. **b**, Stability diagram of the TQD obtained from the charge sensing signal $V_{\text{rf}}$. **c**, Hybrid system of a LD qubit and a ST qubit coupled by the exchange coupling $J^{\text{QQ}}$. **d**, Rabi oscillation of $Q_{\text{LD}}$ (rotation around x-axis) driven by EDSR with $J^{\text{QQ}} \sim 0$ at point RL in Fig. 1b. The data is fitted to oscillations with a Gaussian decay of $T_2^{\text{Rabi}} = 199$ ns. **e**, Pulse sequence used to produce Fig. 1d showing gate voltages $V_{\text{PL}}$ and $V_{\text{PR}}$ applied to the PL and PR gates and a MW burst $V_{\text{MW}}$. **f**, Precession of $Q_{\text{ST}}$ (rotation around z-axis) with a frequency of $f^{ST} = 280$ MHz due to $\Delta E_Z^{ST}$, taken at point E marked by the white circle in (1,1,1) in Fig. 1b, where $J^{QQ}$ and $J^{ST} \sim 0$. The data follow the Gaussian decay with a decay time of 207 ns (see Supplementary Figure 2a) induced by the nuclear field fluctuations[29]. **g**, Pulse sequence used to produce Fig. 1f.

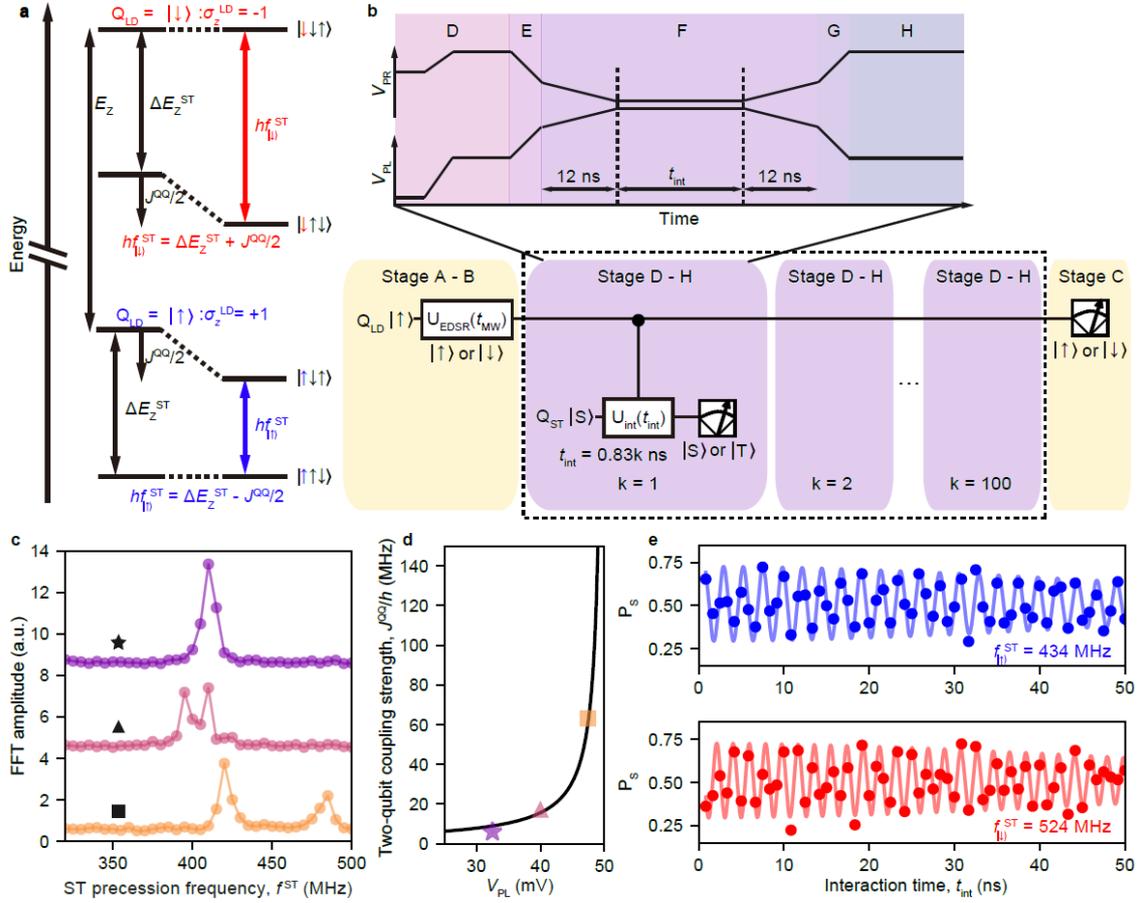

**Figure 2 | ST qubit frequency controlled by the LD qubit.**

**a**, Energy diagram of the two-qubit states for $E_Z \gg \Delta E_Z^{ST}, \Delta E_Z^{QQ} \gg J^{QQ}$ ($J^{ST} = 0$). The ST qubit frequency is equal to $\Delta E_Z^{ST}$ for $J^{QQ} = 0$, and shifts by $\pm J^{QQ}/2$ depending on the $Q_{LD}$ state for finite $J^{QQ}$. **b**, The quantum circuit for demonstrating the phase control of $Q_{ST}$ depending on $Q_{LD}$. After preparing an arbitrary state of $Q_{LD}$ (stages A and B), we run modified stages from D to H (shown in the upper panel) 100 times with $t_{int}$ values ranging from 0.83 to 83 ns to observe the time evolution of $Q_{ST}$ without reinitializing or measuring $Q_{LD}$. Stages A, B and C take 202 μs in total and the part from D to H is 7 μs long. **c**, FFT spectra of $f^{ST}$ with different interaction points shown by the white corresponding symbol in Fig. 1b (traces offset for clarity). In addition to the frequency splitting due to $J^{QQ}$, the center frequency of the two peaks shifts because $\Delta E_Z^{ST}$ is also dependent on the interaction point (Methods). **d**, Interaction point dependence of the ST qubit frequency splitting, i.e. the two-qubit coupling strength $J^{QQ}/h$, fitted with the black model curve (see Supplementary Material, Sec. 4 for the data extraction and fitting). **e**, ST precession for the two $Q_{LD}$ input states $|\uparrow\rangle$ (top panel) and $|\downarrow\rangle$ (bottom panel) fitted with the Gaussian-decaying oscillations with decay times of 72 ns and 75 ns, respectively. The total data acquisition time for each trace is 451 ms.

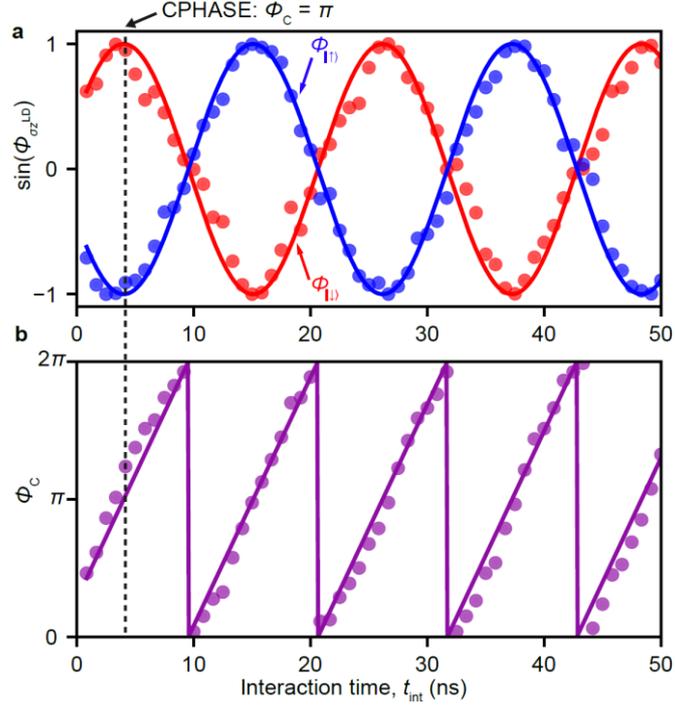

**Figure 3 | Controlled-phase evolution.**

**a**, Interaction time $t_{int}$ dependence of $\phi_{\sigma_z^{LD}}$ controlled by $Q_{LD}$. The blue and red data are for $Q_{LD} = |\uparrow\rangle$ and $|\downarrow\rangle$, respectively. The solid curves are $\sin(\pi J^{QQ}(t_{int} + t_0)/h)$ (red) and $\sin(-\pi J^{QQ}(t_{int} + t_0)/h)$ (blue) where the values of $J^{QQ}$ and $t_0$ are obtained in the MLE. The curves are consistent with the data as expected. **b**, Controlled-phase $\phi_C = \phi_{|\downarrow\rangle} - \phi_{|\uparrow\rangle}$ extracted from Fig. 3a. Including the initial phase accumulated during gate voltage ramps at stage F, $\phi_C$ reaches $\pi$ first at $t_{int} = 4.0$ ns and increases by $\pi$ in every 5.5 ns afterwards.

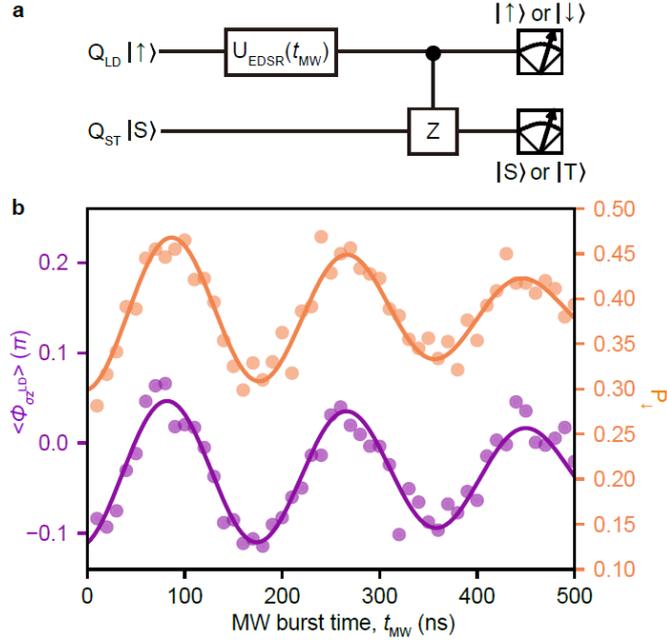

**Figure 4 | Demonstration of the controlled-phase gate for arbitrary control qubit states.**
**a**, The circuit for CPHASE gate demonstration. Here $t_{\text{int}}$ is fixed at 4.2 ns where $\phi_C \approx \pi$ (Methods). **b**, $t_{\text{MW}}$ dependence of the spin-down probability of $Q_{\text{LD}}$, $P_\downarrow$ (yellow) and the averaged $\phi_{\sigma_z^{\text{LD}}}$, $\langle\phi_{\sigma_z^{\text{LD}}}\rangle$ (purple) obtained by the circuit shown in Fig. 4a. $\langle\phi_{\sigma_z^{\text{LD}}}\rangle$ ($= -\pi\langle\sigma_z^{\text{LD}}\rangle/2$) is expected to be proportional to $P_\downarrow$. We see $\langle\phi_{\sigma_z^{\text{LD}}}\rangle$ oscillates depending on the input $Q_{\text{LD}}$ state. The oscillation visibility of $\langle\phi_{\sigma_z^{\text{LD}}}\rangle$ is most probably limited by low preparation fidelity of the input $Q_{\text{LD}}$ state as the visibility of the oscillation in $P_\downarrow$ is also low (see Methods).

**Methods:**

**The device design.**

Our device was fabricated on a GaAs/Al$_{0.3}$Ga$_{0.7}$As heterostructure wafer having a two-dimensional electron gas 100 nm below the surface, grown by molecular beam epitaxy on a semi-insulating (100) GaAs substrate. The electron density $n$ and mobility $\mu$ at a temperature of 4.2 K are $n = 3.21 \times 10^{15}$ m$^{-2}$ and $\mu = 86.5$ m$^2$/Vs in the dark, respectively. We deposited Ti/Au gate electrodes to define the TQD and the charge sensing single electron transistor. A piece of Co metal (micro-magnet, MM) is directly placed on the surface of the wafer to provide a local magnetic field gradient in addition to the external magnetic field applied in-plane (along $z$). The MM geometry is designed based on the numerical simulations of the local magnetic field[23]. The field property is essentially characterized by the two parameters[23]: $dB_x/dz$ at the position of each dot and the difference in $B_z$ between the neighboring dots, $\Delta B_z$ (see Fig. 1a for the definition of the $x$ and $z$ axes). $dB_x/dz$ determines the spin rotation speed by EDSR and is as large as $\sim$ 1 mT/nm[20,23] at the left dot (Extended Data Fig. 1a) allowing fast control of Q$_{LD}$ ($f_{Rabi} > 10$ MHz). At the same time $\Delta B_z$ between the left and center dots, $\Delta B_z^{LC}$, is designed to be $\sim$ 60 mT (Extended Data Fig. 1b) to guarantee the selective EDSR control of Q$_{LD}$ without rotating the spin in the center dot[20,23]. Furthermore, $\Delta B_z$ between the center and right dots, $\Delta B_z^{CR}$, is designed to be $\sim$ 40 mT (Extended Data Fig. 1b) to make the eigenstates of Q$_{ST}$ $|\uparrow\downarrow\rangle$ and $|\downarrow\uparrow\rangle$ rather than $|S\rangle$ and $|T\rangle$ by satisfying $\Delta E_Z^{ST} \gg J^{ST}$. Note that $\Delta E_Z^{ST} = |g|\mu_B \Delta B_z^{CR}$ where $g \sim -0.4$ and $\mu_B$ are the electron $g$-factor and Bohr magneton, respectively. From the design we expect a large variation of $\Delta B_z^{CR}$ when the electron in the center dot is displaced by the electric field. Indeed, we observe a strong influence of the gate voltages on $\Delta B_z^{CR}$, which reaches $\sim$ 100 mT ($\Delta E_Z^{ST}/h \sim 500$ MHz) in the configuration chosen for the two-qubit gate experiment.

**Estimation of the ST precession parameters.**

Here we describe the estimation of the ST precession parameters in Eq. (2) under the influence of a fluctuating single-qubit phase of Q$_{ST}$. Out of the parameters involved, $\phi_{\sigma_z^{LD}}$ is the only parameter assumed to be Q$_{LD}$ state-dependent, and the rest is classified into two types. One is the pulse-cycle-independent parameters, $a, b, J^{QQ}, T_2^*$ and $t_0$ which is constant during the experiment, and the other is the pulse-cycle-dependent parameters, $\sigma_z^{LD}, \Delta E_Z^{ST}$ and $\phi_0$, which can change cycle by cycle. Each pulse cycle consists of pulse stages from A to C as shown in Fig. 2b. We run the pulse cycle consecutively with a MW frequency fixed at 17.26 GHz and collect the data while Q$_{LD}$ drifts between on- and off-resonances with the MW burst due to the nuclear field fluctuation. To decrease the uncertainty of the estimated parameters, we choose the cycles during which the spin flip of Q$_{LD}$ is unlikely in the following manner. The cycles throughout which Q$_{LD}$ is likely to be $|\downarrow\rangle$ are post-selected by the condition that Q$_{LD}$ is on-resonance (i.e., Rabi oscillation of Q$_{LD}$ is observed in ensemble-averaged data from nearby cycles) and the final state of Q$_{LD}$ is measured to be $|\downarrow\rangle$ at pulse stage C. Similarly, the cycles for Q$_{LD}$ = $|\uparrow\rangle$ are post-selected by the condition that Q$_{LD}$ is off-resonance and the final state of Q$_{LD}$ is measured to be $|\uparrow\rangle$. The data structure and the index definitions for MLE are summarized in Extended Data Table 1. $k$ is the index of the interaction time such that $t_{int} = 0.83 \times k$ ns with $k$ ranging from 1 to 100. $m$ is the pulse cycle index ranging from 1 (2001) to 2000 (4000) for Q$_{LD}$ prepared in $|\uparrow\rangle$ ($|\downarrow\rangle$). The estimation procedure is the following. From all the readout results of Q$_{ST}$ (stage H) obtained in the cycles, we first estimate the five pulse-cycle-independent parameters by MLE. Note that $J^{QQ}$ may have a small pulse-cycle-dependent component due to charge noise but this effect is captured as additional fluctuation in $\Delta E_Z^{ST}$ and $\phi_0$ in our model. We apply MLE to $100 \times 4000$ readout results of Q$_{ST}$, $r_m^k = 1$ (0) for Q$_{ST}$ = $|S\rangle$ ($|T\rangle$). To this end, we first introduce the likelihood P$_m$ defined in the eight dimensional parameter space as

$$P_m(a, b, t_0, J^{QQ}, T_2^*, \sigma_z^{LD}, \phi_0, \Delta E_Z^{ST}) = \prod_{k=1}^{100}(r_m^k P_{S,model} + (1-r_m^k)(1-P_{S,model})) \quad (3)$$

where P$_{S,model}$ is defined in Eq. (2). We calculate P$_m$ on a discretized space within a chosen parameter range (Extended Data Table 2) using a single cycle data. Then we obtain P$_m$ for the target five parameters as a marginal distribution by tracing out the pulse-cycle-dependent parameters,

$$P_m(a, b, t_0, J^{QQ}, T_2^*) = \sum_{\sigma_z^{LD}} \sum_{\phi_0} \sum_{\Delta E_Z^{ST}} P_m(a, b, t_0, J^{QQ}, T_2^*, \sigma_z^{LD}, \phi_0, \Delta E_Z^{ST}). \quad (4)$$

Repeating this process for all pulse cycles, we obtain the likelihood P as

$$P(a, b, t_0, J^{QQ}, T_2^*) = \prod_m P_m(a, b, t_0, J^{QQ}, T_2^*). \quad (5)$$

We choose the maximum of P as the estimator for $a, b, t_0, J^{QQ}$ and $T_2^*$, obtaining $a = 0.218 \pm 0.005$, $b = 0.511 \pm 0.003$, $t_0 = 1.53 \pm 0.17$ ns, $J^{QQ}/h = 90.2 \pm 0.3$ MHz, $T_2^* = 211 \pm 37$ ns.

Once these values are fixed, we estimate the pulse-cycle-dependent parameters, $\sigma_z^{LD}, \phi_0$ and $\Delta E_Z^{ST}$, for each cycle $m$. Note that $\sigma_z^{LD}$ could be prepared deterministically if the state preparation of Q$_{LD}$ were ideal, but here we treat it as one of the parameters to be estimated because of a finite error in the Q$_{LD}$ state preparation. We again evaluate the likelihood $P_m(\sigma_z^{LD}, \phi_0, \Delta E_Z^{ST})$ defined in a discretized three dimensional space of its parameters using Eq. (3) and find their values that maximize the likelihood.

Based on the values of $a, b, T_2^*$ and $\phi^{ST}$ determined above, we can directly estimate $\phi_{\sigma_z^{LD}}$ controlled by Q$_{LD}$ for each $t_{int}$ without presumptions on the value of $J^{QQ}$. To this end, we search for the parameter $\phi_{\sigma_z^{LD}}$ that maximizes the likelihood

$$P^k(\phi_{\sigma_z^{LD}}) = \prod_m (r_m^k P_{S,\text{model}} + (1 - r_m^k)(1 - P_{S,\text{model}})). \quad (6)$$

The obtained estimators for $\phi_{|\downarrow\rangle}$ and $\phi_{|\uparrow\rangle}$ are consistent with the expected values $\pm \pi J^{QQ}(t_{\text{int}} + t_0)/h$ calculated from $J^{QQ}/h$ and $t_0$ found above (see Fig. 3a).

The ensemble-averaged phase $\langle \phi_{\sigma_z^{LD}} \rangle$ is obtained based on a similar estimation protocol. Here we estimate $\phi_{\sigma_z^{LD}}$ for each $m$ with fixed $k = 5$ ($t_{\text{int}} = 4.2$ ns) to yield $\phi_C \approx \pi$ from the likelihood $P_m^{k=5} = r_m^{k=5} P_{S,\text{model}} + (1 - r_m^{k=5})(1 - P_{S,\text{model}})$ and then take the average of the estimated values for 800 pulse cycles. The oscillation visibility of $\langle \phi_{\sigma_z^{LD}} \rangle$ in Fig. 4 is limited by three factors, low preparation fidelity of the input $Q_{LD}$ state, estimation error of $\phi_{\sigma_z^{LD}}$ and CPHASE gate error. The first contribution is likely to be dominant as the visibility of the oscillation in $P_\downarrow$ is correspondingly low. Note that the effect of those errors is not visible in Fig. 3 because the most likely values of $\phi_{\sigma_z^{LD}}$ are plotted.

**Data availability**

The data that support the findings of this study are available from the corresponding authors upon reasonable request.

**References:**

[37]We used MATHEMATICA RADIA package available at http://www.esrf.fr/ for MM field calculations.

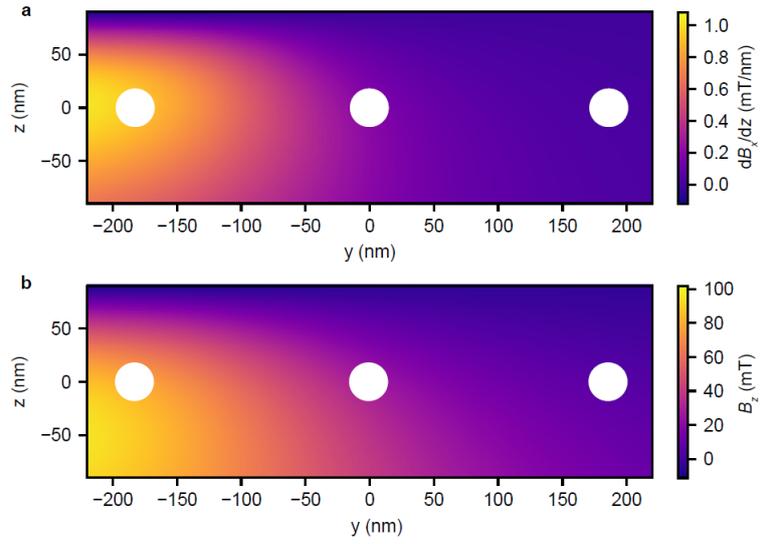

**Extended Data Figure 1 | Local magnetic field simulation.**

**a**, Simulated distribution of the slanting field $dB_x/dz$ created by a MM for the design shown in Fig. 1a calculated by the boundary integral method[37]. The white circles indicate the positions of the three dots from the device lithography design. **b**, Simulated local Zeeman field $B_z$.

| $Q_{LD}$ | Interaction time index $k$ | $k = 1$ | ... | $k = 100$ |
| | Pulse cycle index $m$ | ($t_{\text{int}} = 0.83$ ns) | ... | ($t_{\text{int}} = 83$ ns) |
|---|---|---|---|---|
| | $m = 1$ | $r_{m=1}^{k=1}$ | ... | $r_{m=1}^{k=100}$ |
| $\sigma_z^{LD} = \|\uparrow\rangle$ | ⋮ | ⋮ | ... | ⋮ |
| | $m = 2000$ | $r_{m=2000}^{k=1}$ | ... | $r_{m=2000}^{k=100}$ |
| | $m = 2001$ | $r_{m=2001}^{k=1}$ | ... | $r_{m=2001}^{k=100}$ |
| $\sigma_z^{LD} = \|\downarrow\rangle$ | ⋮ | ⋮ | ... | ⋮ |
| | $m = 4000$ | $r_{m=4000}^{k=1}$ | ... | $r_{m=4000}^{k=100}$ |

**Extended Data Table 1 | Data structure for the ST precession parameter estimation.**

Collected data set for the ST precession parameter estimation. $m$, $k$ and $r_m^k$ reperesent the pulse cycle index, the interaction time index $t_{\text{int}}(k) = 0.83k$ ns and the readout result of $Q_{ST}$: $r_m^k = 1$ (0) for $Q_{ST} = |S\rangle$ ($|T\rangle$). For each interaction time and $Q_{LD}$ state, we have 2000 readout results whereas for each pulse cycle, we have 100 readout results of $Q_{ST}$.

| Parameter | Minimum | Maximum | Numbre of discretized points | Pulse-cycle-dependency |
|---|---|---|---|---|
| $a$ | 0.208 | 0.223 | 16 | |
| $b$ | 0.501 | 0.516 | 16 | |
| $J^{QQ}/h$ (MHz) | 89.8 | 91.3 | 16 | No |
| $t_0$ (ns) | 1.4 | 1.7 | 16 | |
| $T_2^*$ (ns) | 160 | 310 | 16 | |
| $\sigma_z^{LD}$ | $-1$ | 1 | 2 | |
| $\phi_0$ | $\pi/16$ | $2\pi$ | 32 | Yes |
| $\Delta E_Z^{ST}/h$ (MHz) | 465 | 496 | 32 | |

**Extended Data Table 2 | Parameter space for Eq. (3).**

The eight dimensional parameter space for evaluating Eq. (3). The search ranges of the five pulse-cycle-independent parameters are determined based on prior, coarse estimation results over wide spans.

**Supplementary Information:**

**A fast quantum interface between different spin qubit encodings**

A. Noiri, T. Nakajima, J. Yoneda, M. R. Delbecq, P. Stano, T. Otsuka, K. Takeda, S. Amaha, G. Allison, K. Kawasaki, Y. Kojima, A. Ludwig, A. D. Wieck, D. Loss, and S. Tarucha

**1. The Zeeman energy difference between the nighboring dots.**

To confirm that the oscillation observed in Fig. 1f is driven by $\Delta E_Z^{ST}$, we measure the EDSR spectrum of the center and right dots. We prepare the doubly-occupied singlet state in the right dot and transfer one of the electrons to the center dot adiabatically with respect to $\Delta E_Z^{ST}$ to initialize the two spins to $|\uparrow\downarrow\rangle$[26]. Then we measure an EDSR spectrum by frequency chirping with a depth of 10 MHz and MW burst length of 20 μs[S1]. Here the spectrum is taken at the same gate voltage configuration as the one used for taking data in Fig. 1f. Then the two-spin state is measured by Pauli spin blockade[26]. When one of the spins rotates and the two-spin state becomes either one of the polarized triplets, the two electrons remain blocked from returning to the doubly-occupied singlet. Fig. S1a shows the MW frequency $f_{MW}$ dependence of the singlet-return probability $P_S$ with $B_{ext} = 3.105$ T. The two dips around $f_{MW} \sim 16.1$ GHz and 16.4 GHz correspond to the resonances of the center and right dots, respectively, which are separated by ~300 MHz in agreement with the ST precession frequency $f^{ST}$ in Fig. 1f. Because of the strong gate-voltage dependence of $\Delta E_Z^{ST}$ (see Methods Sec. 1) the obtained $f^{ST}$ in this measurement is much smaller than that measured in the two-qubit gate experiment ($\Delta E_Z^{ST}/h = (f_{|\uparrow\rangle}^{ST} + f_{|\downarrow\rangle}^{ST})/2 \sim 480$ MHz, see Fig. 2e).

We also measure the EDSR spectrum of the left and center dots with $B_{ext} = 3.15$ T as shown in Fig. S1b. The two dips of $P_S$ around $f_{MW} \sim 17.1$ GHz and 17.6 GHz correspond to the resonances of the center and left dots, respectively. As a consequence, we obtain $\Delta E_Z^{QQ}/h \sim$500 MHz.

**2. Full control of Q$_{ST}$**

A set of universal quantum gates in a two-qubit system can be constructed by a CPHASE gate and arbitrary single-qubit gates for each qubit. Arbitrary single-qubit gate operations for Q$_{LD}$ can be realized by EDSR (see Fig. 1d). In this section we demonstrate full control of Q$_{ST}$ for completeness. In the following experiment, we quench $J^{QQ}$ to decouple Q$_{ST}$ from Q$_{LD}$.

The rotation of Q$_{ST}$ around *z*-axis and *x*-axis is mediated by $\Delta E_Z^{ST}$ and $J^{ST}$, respectively, as shown in Fig. 1c. Figure S2a shows the precession around *z*-axis which is measured by initializing Q$_{ST}$ to $|S\rangle$ and projecting the final state along *x*-axis ($|S\rangle$ or $|T\rangle$) using a pulse sequence shown in Fig. 1g. During the evolution in stage F, $J^{ST}$ is quenched ($\Delta E_Z^{ST} \gg J^{ST}$) as the evolution point is far detuned from the resonance of (1,1,1) and (1,0,2) charge states. The fit with the Gaussian decaying oscillation curve gives $f^{ST} = 280$ MHz, which is consistent with the FFT spectrum in Fig. S2b. Here, the precession visibility is mainly influenced by the error during the state preparation and measurement. The dominant error source is likely the incomplete nonadiabaticity during the state transfers in pulse stages E (for state preparation) and G (for measurement) [the error in charge state discrimination is negligible as described in the main text]. To keep the adiabaticity with respect to the inter-dot tunnel coupling, we choose the pulse rise/fall time of 5 ns, which is too slow to switch $J^{ST}$ nonadiabatically against $\Delta E_Z^{ST}/h \sim 280$ MHz. As a result, the prepared state in stage E is inclined to $|\uparrow\downarrow\rangle$ from $|S\rangle$, decreasing the ST precession visibility even though the system is coherent. This error could be suppressed by, instead, initializing to $|\uparrow\downarrow\rangle$ and subsequently rotating around y-axis[30,33], although we do not experimentally pursue this alternative preparation. We also note that the ST precession visibility can decrease due to the state leakage to non-qubit states during turning on and off $J^{QQ}$. We avoid this problem by adiabatically turning on and off $J^{QQ}$ with respect to $\Delta E_Z^{QQ}$ (see Supplementary Material Sec. 3) and obtain a similar visibility of the ST precessions measured in $J^{QQ}$ on (Fig. 2e) and off (Fig. 1f).

Next we demonstrate the qubit control around x-axis using a pulse sequence shown in Fig. S2d. Here Q$_{ST}$ is initialized to $|\uparrow\downarrow\rangle$ by slow adiabatic passage, kept at a detuned point (stage F) and projected along z-axis (($|\uparrow\downarrow\rangle$ or $|\downarrow\uparrow\rangle$)) using the reverse process of the initialization[11,26]. Figure S2e shows the evolution of Q$_{ST}$ as a function of the detuned point. As the point approaches the resonance of (1,1,1) and (1,0,2), $J^{ST}$ increases and the rotation axis is inclined toward the x direction, eventually realizing the rotation around the x-axis for $\Delta E_Z^{ST} \ll J^{ST}$. In this scheme, however, the dephasing time decreases as $J^{ST}$ is increased due to the enhanced exchange noise[S2]. To improve the control quality, we employ resonantly driven rotation of Q$_{ST}$ around the x-axis in the rotating frame[30,33]. Here we choose a detuned point where $\Delta E_Z^{ST} > J^{ST}$ is satisfied and modulate $J^{ST}$ at the qubit resonance frequency, $\sqrt{J^{ST^2} + \Delta E_Z^{ST^2}}/h$, by applying a MW burst to PC gate (see Fig. 1a). Then Q$_{ST}$ exhibits the Rabi rotation at a frequency proportional to the modulation amplitude. Figure S2g shows the MW frequency dependence of the Rabi oscillations. We find the resonance at $f_{MW}^{ST} = 345$ MHz, where we see a clear coherent oscillation as shown in Fig. S2h. These results demonstrate full control of Q$_{ST}$ and therefore, our system is capable of universal two-qubit manipulations.

**3. The origin of the two-qubit gate.**

In this section we discuss the conditions for Eq. (1) to be a good approximation[18]. The general Hamiltonian of the three-spin system in the (1,1,1) charge state under a magnetic field is given by

$$\mathcal{H} = \mathcal{H}_Z + \mathcal{H}_J \quad (S1)$$

$$\mathcal{H}_Z = -E_Z \hat{\sigma}_z^{LD}/2 - (\Delta E_Z^{QQ} + E_Z)\hat{\sigma}_z^C/2 - (\Delta E_Z^{ST} + \Delta E_Z^{QQ} + E_Z)\hat{\sigma}_z^R/2 \quad (S2)$$

$$\mathcal{H}_J = J^{QQ}(\hat{\boldsymbol{\sigma}}^{LD} \cdot \hat{\boldsymbol{\sigma}}^C - \mathbf{1})/4 + J^{ST}(\hat{\boldsymbol{\sigma}}^C \cdot \hat{\boldsymbol{\sigma}}^R - \mathbf{1})/4 \quad (S3)$$

where $\mathcal{H}_Z$ and $\mathcal{H}_J$ represent the Zeeman energy term and the exchange coupling, respectively. Here $\hat{\boldsymbol{\sigma}}^{LD}$ ($\hat{\sigma}_z^{LD}$), $\hat{\boldsymbol{\sigma}}^C$ ($\hat{\sigma}_z^C$) and $\hat{\boldsymbol{\sigma}}^R$ ($\hat{\sigma}_z^R$) are the Pauli operators (and their *z*-component) of the spin in the left, center and right dots, respectively. $E_Z$ is the Zeeman energy of the spin in the left dot. $\Delta E_Z^{QQ}$ and $\Delta E_Z^{ST}$ are the Zeeman energy difference between the left and center dots and between the right and center dots, respectively. We assume exchange couplings only between neighboring dots, $J^{QQ}$ ($J^{ST}$) between the left (right) and center dots. The Zeeman energy difference between the neighboring dots $\Delta E_Z^{QQ}$ ($\Delta E_Z^{ST}$) competes with the exchange coupling $J^{QQ}$ ($J^{ST}$). In the case of $\Delta E_Z^{QQ}, \Delta E_Z^{ST} \gg J^{QQ}, J^{ST}$ which is the case in our experiments, the three-spin eigenstates are well approximated by three isolated spins $|\sigma_z^{LD}\sigma_z^C\sigma_z^R\rangle$ rather than the exchange dominated states of the doublets and quardruplets[S3]. Those three-spin states, mainly determined by $\mathcal{H}_Z$, are perturbed by $\mathcal{H}_J$. At the two-qubit interaction point which is near the resonance of (1,1,1) and (2,0,1) charge state, $J^{QQ} \gg J^{ST} \sim 0$ is satisfied and therefore the second term of Eq. (S3) is neglected. Furthermore, we apply large external magnetic field such that $E_Z \gg \Delta E_Z^{QQ}, \Delta E_Z^{ST}$ to energetically separate $|\uparrow\uparrow\uparrow\rangle$ and $|\downarrow\downarrow\downarrow\rangle$ from the two-qubit subspace of $|\uparrow\uparrow\downarrow\rangle, |\uparrow\downarrow\uparrow\rangle, |\downarrow\uparrow\downarrow\rangle$ and $|\downarrow\downarrow\uparrow\rangle$, with the consequence that the leakage from the qubit states to those fully spin-polarized states becomes negligible. The state leakage to the other non-qubit states, $|\downarrow\uparrow\uparrow\rangle$ and $|\uparrow\downarrow\downarrow\rangle$, is also suppressed by adiabatically turning on and off $J^{QQ}$ with respect to $\Delta E_Z^{QQ}$[28]. Therefore we can restrict ourselves only to the two-qubit subspace under the condition of $E_Z \gg \Delta E_Z^{QQ}, \Delta E_Z^{ST} \gg J^{QQ} \gg J^{ST}$. Here the first, second and third terms of Eq. (S2) are simplified to the first and the second terms of Eq. (1). In addition, the first term of Eq. (S3) can be approximated by the last term of Eq. (1). In our experiment, the relevant parameter values at the two-qubit interaction configuration are $E_Z/h \sim 17$ GHz, $\Delta E_Z^{QQ}/h \sim \Delta E_Z^{ST}/h \sim 0.5$ GHz, $J^{QQ}/h \sim 0.09$ GHz and $J^{ST}/h \sim 0$ GHz and therefore the above conditions are satisfied and Eq. (1) is a good approximation.

### 4. The controllability of $J^{QQ}$ by the gate voltages.

In this section we discuss the three-spin state energy diagram in our experimental setup and its gate voltage dependence. Fig. S3a shows the calculated energy levels as a function of the gate voltages $V_{PL}$ and $V_{PR}$ which are changed to detune the energies between the outer dots while keeping the center dot energy level fixed[26,S3]. Here we assume a common inter-dot tunnel coupling $t$ between neighboring dots and Zeeman energies $E_Z/h = 17$ GHz, $\Delta E_Z^{QQ}/h = \Delta E_Z^{ST}/h = 0.5$ GHz. The red and blue curves in Fig. S3c show the Q$_{LD}$ state dependent ST precession frequencies $f_{\sigma_z^{LD}}^{ST}$ as a function of $V_{PL}$. Figure S3b shows the $f_{\sigma_z^{LD}}^{ST}$ spectra measured at the three interaction points marked by the star, triangle and square symbols in Fig. 1b. Here we obtain the spectra by Bayesian estimation[29,30] (see also Methods) instead of the FFT used in Fig. 2c. By fitting the spectra, we obtain splittings of $f^{ST}_{|\downarrow\rangle}$ and $f^{ST}_{|\uparrow\rangle}$, $f^{ST}_{|\downarrow\rangle} - f^{ST}_{|\uparrow\rangle} = 6.08 \pm 0.01, 17.18 \pm 0.01$ and $63.13 \pm 0.01$ MHz for each point. These values are plotted in Fig. 2d. Then we fit the obtained gate voltage dependence of $f^{ST}_{|\downarrow\rangle} - f^{ST}_{|\uparrow\rangle}$ with the model curve calculated from Fig. S3a as shown by the black curve. We find an agreement between the theory model and the data upon choosing $t = 1.2$ GHz and the lever arm of the detuning energy against $V_{PL}$ of 18 meV/V.

## 5. The model of the ST precession under $J^{QQ}$.

Here we explain the ST precession model $P_{S,model}$ in Eq. (2). We assume that $P_{S,model}$ is a Gaussian decaying oscillation function[S4] with an oscillation frequency $f^{ST}_{\sigma^{LD}_z}$. The decay time $T_2^*$ is the dephasing time of $Q_{ST}$ determined by fluctuating $\phi^{ST}$ (due to nuclear field and charge noise) occuring within a single pulse cycle of 700 μs (in the black dashed square in Fig. 2b). The oscillations amplitude $a = 0.218$ and the mean value of the oscillation $b = 0.511$ are determined by three factors, i.e., initialization error of $Q_{ST}$, tilt of the precession axis determined by $J^{ST}/\Delta E^{ST}_Z$ and the readout error of $Q_{ST}$. The phase of $Q_{ST}$ is described by two terms, $Q_{LD}$-controlled phase $\phi_{\sigma^{LD}_z} = -\pi\sigma^{LD}_z J^{QQ}(t_{int} + t_0)/h$ and the single-qubit phase $\phi^{ST} = 2\pi\Delta E^{ST}_Z(t_{int} + t_{ramp})/h + \phi_0$ which accumulates independently from $Q_{LD}$. Here $t_0$ accounts for the effective total time for turning on and off $J^{QQ}$ with voltage pulses. The estimated value of $t_0 = 1.53$ ns found by MLE is much smaller than the voltage pulse ramp time of $t_{ramp} = 24$ ns as $J^{QQ}$ rapidly increases only near the resonance of (1,1,1) and (2,0,1) charge states (see Fig. 2d). Since $\Delta E^{ST}_Z$ also varies during the voltage ramps due to the inhomogeneity of the MM-induced magnetic field, an additional phase $\phi_0$ is necessary to properly describe $\phi^{ST}$. We note the fluctuation of $\phi_0$ is independent from that of $\Delta E^{ST}_Z$ due to the inhomogeneity of the nuclear Overhauser field. In this model, any possible fluctuation of $J^{QQ}$ due to charge noise is equivalent to an additional fluctuation in $\Delta E^{ST}_Z$ and $\phi_0$, and therefore does not have to be parametrized separately.

## 6. Coherence of the two qubits.

Here we discuss possible limiting factors of $T_2^*$ in Eq. (2) following the analysis presented in Ref. 29. $T_2^*$ is the dephasing time caused by the two fluctuators: the nuclear field and charge noise, within the data acquisition time. To evaluate the characteristic time scale of the fluctuation, we calculate the time correlator of $f^{ST}_{\sigma^{LD}_z}$, $C_f(\Delta t) = f^{ST}_{\sigma^{LD}_z}(t + \Delta t) - f^{ST}_{\sigma^{LD}_z}(t)$ as a function of time delay, $\Delta t$ (see Ref. 29). The histogram of the correlator has a main peak at $C_f = 0$ MHz and two side peaks at $\pm J^{QQ}/h = \pm 90$ MHz due to flip-flops of the $Q_{LD}$ state (Fig. S4a). All of the peaks show Gaussian distribution and their variance $\sigma_f^2(\Delta t)$, which has similar values between each peak, determines $T_2^*$ as $T_2^*(\Delta t) \propto 1/\sigma_f(\Delta t)$. Fig. S4b shows the $\Delta t$ dependence of $\sigma_f^2$ calculated from the main peak of $C_f$. We identify two different regimes, 50 ms $< \Delta t$ and $\Delta t <$ 50 ms. For 50 ms $< \Delta t$, the variance shows the dependence $\sigma_f^2 \propto \Delta t^{0.8}$ similar to Ref. 29, where the diffusion of nuclear spins is suggested to be the origin. In this regime, the variance and thereby $T_2^*$ are limited by the Overhauser field fluctuation. On the other hand, in $\Delta t < 50$ ms, the variance displays saturation. This suggests another noise source with a larger high-frequency tail, which we believe is due to the charge noise becoming dominant. Therefore we suppose that in our two-qubit gate experiment, the coherence of $Q_{ST}$ is limited by charge noise. In this regime, the observed dephasing time dominates the dephasing of the whole two-qubit system, and therefore we conclude that the dephasing time of our CPHASE gate is 211 ns.

In contrast, for the ST precession measurement data shown in Fig. 2e, a much longer data acquisition time of 451 ms results in $T_2^* \approx 70$ ns, limited by the nuclear spins.

## 7. The unconditional phase accumulation during the CPHASE gate.

Here we describe the condition where the CPHASE gate including the single-qubit phase gates of $Q_{ST}$ and $Q_{LD}$ is realized[S5,18]. From Eqs. (1) and (2), the two-qubit interaction in the basis states of $|\uparrow\uparrow\downarrow\rangle, |\uparrow\downarrow\uparrow\rangle, |\downarrow\uparrow\downarrow\rangle$ and $|\downarrow\downarrow\uparrow\rangle$ is given by the unitary operation,

$$U(t_{\text{int}}) = Z^{\text{LD}}(-\phi^{\text{LD}})Z^{\text{ST}}(-\phi^{\text{ST}})\begin{pmatrix} 1 & 0 & 0 & 0 \\ 0 & e^{-i\phi_{|\uparrow\rangle}} & 0 & 0 \\ 0 & 0 & e^{i\phi_{|\downarrow\rangle}} & 0 \\ 0 & 0 & 0 & 1 \end{pmatrix}$$

$$= Z^{\text{LD}}(-\phi^{\text{LD}} + \phi_{|\downarrow\rangle})Z^{\text{ST}}(-\phi^{\text{ST}} - \phi_{|\uparrow\rangle})\begin{pmatrix} 1 & 0 & 0 & 0 \\ 0 & 1 & 0 & 0 \\ 0 & 0 & 1 & 0 \\ 0 & 0 & 0 & e^{-i(\phi_{|\downarrow\rangle} - \phi_{|\uparrow\rangle})} \end{pmatrix} \quad (S4)$$

where $Z^{\text{LD}}(\phi)$ and $Z^{\text{ST}}(\phi)$ represent qubit rotation around the z-axis by angle $\phi$ for $Q_{LD}$ and $Q_{ST}$, respectively. Up to single-qubit phases (neglecting $Z^{\text{LD}}(-\phi^{\text{LD}} + \phi_{|\downarrow\rangle})$ and $Z^{\text{ST}}(-\phi^{\text{ST}} - \phi_{|\uparrow\rangle})$), the CPHASE gate is realized for any interger $n$ such that $\phi_C = \phi_{|\downarrow\rangle} - \phi_{|\uparrow\rangle} = 2\pi J^{QQ}(t_{\text{int}} + t_0)/h = \pi(2n + 1)$. This condition is met in our experiment with $t_{\text{int}} = 4.0 + 11n$ ns where 4.0 ns comes from the initial phase due to a finite value for $t_0$.

More strictly, to construct the CPHASE gate without ignoring the single-qubit phases, each single-qubit phase should satisfy $\phi^{\text{LD}} - \phi_{|\downarrow\rangle} = 2\pi l$ and $\phi^{\text{ST}} + \phi_{|\uparrow\rangle} = 2\pi m$ where $l$ and $m$ are integers. This can be realized by a phase gate of $Q_{ST}$ and $Q_{LD}$ or a decoupled CPHASE gate[7,14], although we do not perform them in our experiment.

## 8. The quality factor of the CPHASE gate.

Here we discuss the performance of the CPHASE gate. A convenient figure of merit is the quality factor $Q$ representing the number of possible CPHASE gate operations within the dephasing time. It is evaluated to be $Q = 2J^{QQ}T_2^*/h = 38$ in our experiment. For comparison, Ref. 7 uses a two-qubit gate with $Q \sim 10$ and evaluates the lower bound of the gate fidelity to be 85 % extracted from the Bell state tomography. Our value of $Q$ therefore suggests that our CPHASE gate fidelity could be potentially higher than 85 %. However, the gate fidelity has to be characterized by more elaborate approaches such as process tomography or randomized bechmarking in future experiments. Si-based devices with much better single-qubit gate fidelity will allow these detailed characterizations.

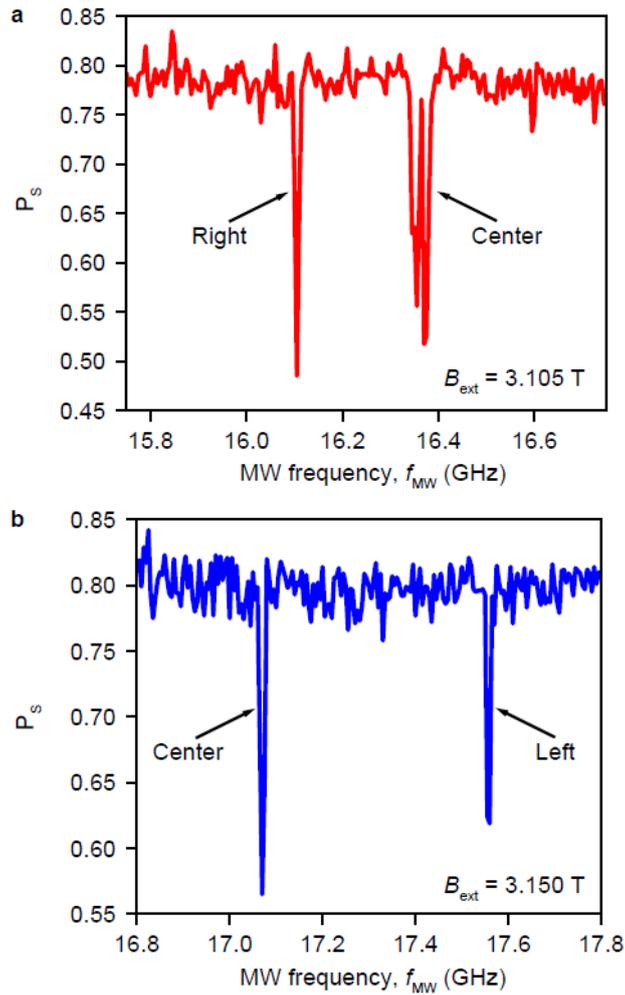

**Supplementary Figure 1 | EDSR spectrum measured between the neighboring dots.**
**a**, EDSR spectrum for the center and right dots taken at the same configuration as the one to obtain the data shown in Fig. 1f. The resonance frequency of the center dot is higher than that of the right dot as the local Zeeman field of the center dot induced by the micro-magnet is larger than that of the right dot (Extended Data Fig. 1b). The separation of the resonance condition ($\sim 300$ MHz) is consistent with the ST precession frequency $f^{ST}$ in Fig. 1f. **b**, EDSR spectrum for the left and center dots. The resonance frequencies are separated by $\sim 500$ MHz.

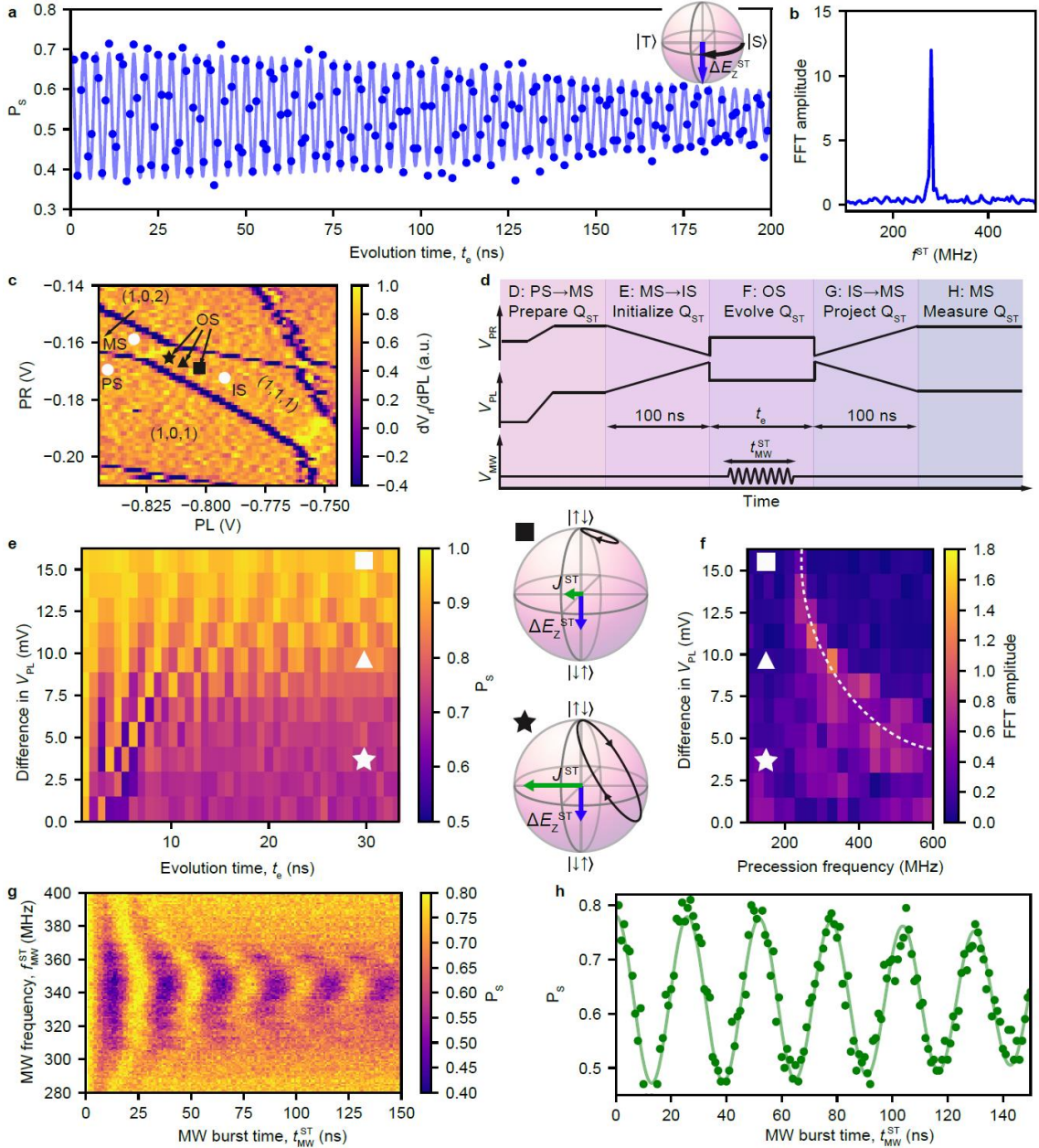

**Supplementary Figure 2 | Full control of Q$_{ST}$.**

**a**, Precession around the *z*-axis due to $\Delta E_Z^{ST}$ by quenching $J^{ST}$. The data and the fitting curve is the same as the one shown in Fig. 1f. The fitting gives the dephasing time of $207 \pm 11$ ns. **b**, FFT spectrum of Fig. S2a showing a peak at $f^{ST} = 280$ MHz in agreement with the fitting (see Fig. 1f). **c**, Stability diagram of the TQD used for demonstrating the full control of Q$_{ST}$. [The gate voltage configuration differs from Fig. 1b and the data were taken in a different cool-downs]. **d**, Pulse sequence used to demonstrate the control of Q$_{ST}$ around *x*-axis by $J^{ST}$. In stages E and G, $|\uparrow\downarrow\rangle$ and $|S\rangle$ ($|\downarrow\uparrow\rangle$ and $|T\rangle$) are interconverted adiabatically, allowing a *z*-axis readout of Q$_{ST}$ by the Pauli spin blockade[13]. For taking the data in **g**, a MW burst with frequency $f_{MW}^{ST}$ and duration $t_{MW}^{ST}$ is applied to the PC gate at stage F. **e**, The state evolution during the $Q_{ST}$ rotation. The trajectories of $Q_{ST}$ at detuned points marked by the square and star symbols are illustrated in the laboratory-frame Bloch spheres shown on the right. **f**, FFT spectra of the data in **e**. The white dashed curve is an eye guide of the spectral peaks drawn at $\sqrt{J^{ST2} + \Delta E_Z^{ST2}}/h$. **g**, A MW frequency dependence of resonantly driven rotation, taken at the point shown by the black triangle in Fig. S2c. **h**, Rabi oscillation of $Q_{ST}$ measured at the resonance, $f_{MW}^{ST} = 345$ MHz.

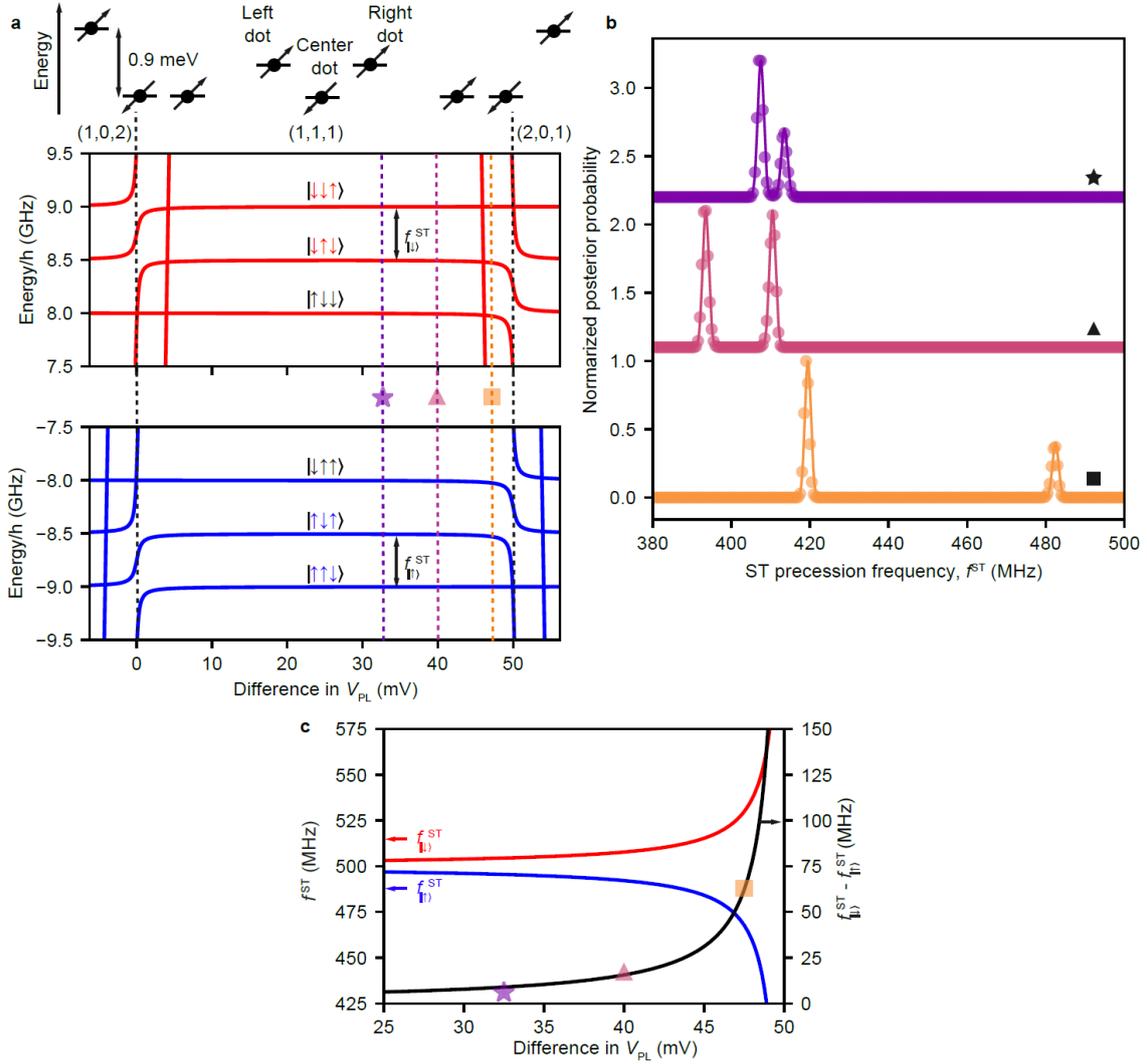

**Supplementary Figure 3 | Three-spin state energy diagram.**

**a**, Three-spin state energy diagram as a function of $V_{PL}$ with a corresponding energy level of each dot. The voltages are plotted as difference from the degeneracy of (1,1,1) and (1,0,2) charge states, at $(V_{PL}, V_{PR}) = $ (-348 mV, -208 mV) (see Fig. 1b). The degeneracy between (1,1,1) and (2,0,1) charge states is located at $(V_{PL}, V_{PR}) = $ (-298 mV, -248 mV). We assume that the energy difference between (1,0,2) and (2,0,1) charge states, corresponding to the difference in $(V_{PL}, V_{PR})$ being (50 mV, -40 mV), is 0.9 meV. This gives an agreement with the measured data (three symbols) in Fig. 2d. The positions of the interaction points used in Figs. 2c and 2d are shown by dashed lines with the corresponding symbols. **b**, Posterior probability of $f^{ST}$ calculated by Bayesian estimation from the data used in Fig. 2c (traces offset for clarity). Each trace is normarized by its maximum. The solid curves are the fit with the weighted sum of two Gaussian distributions. **c**, $f^{ST}_{\sigma_z^{LD}}$ (the red and blue curves) and $f^{ST}_{|\downarrow\rangle} - f^{ST}_{|\uparrow\rangle}$ (the black curve) extracted from **a** as a function of $V_{PL}$. The three symbols are obtained from **b**.

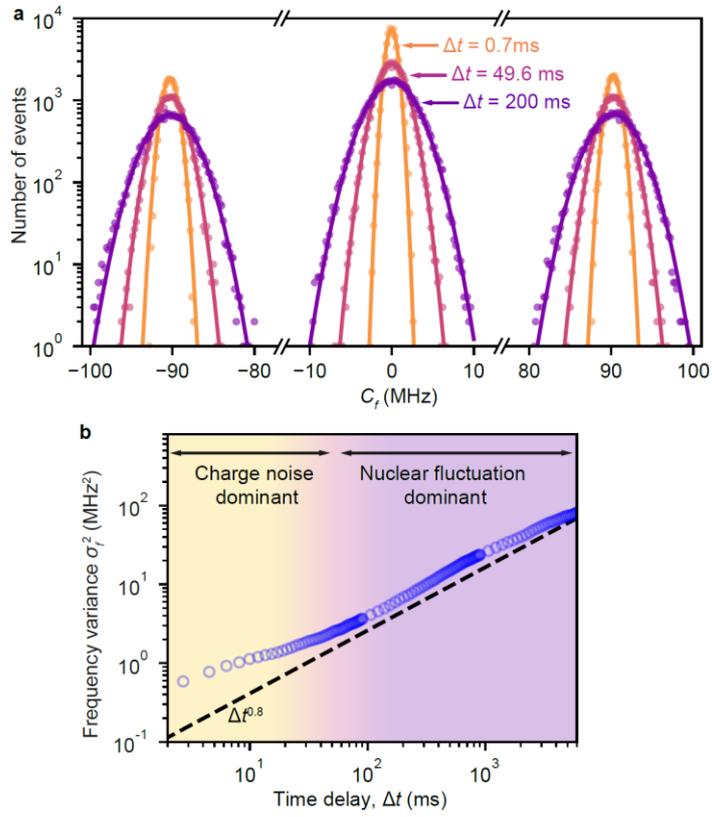

**Supplementary Figure 4 | Variance of the ST precession frequency.**

**a**, Histogram of the time correlator of the estimated ST precession frequency having a main peak at zero frequency and two side peaks at $\pm J^{QQ}/h$. Each peak fits well to a Gaussian distribution. **b**, $\Delta t$ dependence of the estimated ST precession frequency variance calculated from the main peak of $C_f$ (blue open circles). The dotted line is $\sigma_f^2 \propto \Delta t^{0.8}$ reflecting the nuclear spin diffusion[29].